\documentclass[ms,11pt]{egs}
  \usepackage{times}                   

\usepackage{graphicx}

\begin{document}

\title{Extremum Statistics: a Framework for Data Analysis.}  
\author[1]{S. C. Chapman,
G. Rowlands}
\affil[1]{
Physics Dept. Univ. of Warwick,
Coventry CV4 7AL, UK
}
\author[2]{
N. W. Watkins}
\affil[2]{British Antarctic Survey, High 
Cross, Madingley Rd., Cambridge CB3 0ET, UK}
\date{\today}
\journal{\NPG}
\msnumber{12345}

\maketitle
\begin{abstract}
Recent work has suggested that in highly
correlated systems, such as sandpiles, turbulent fluids, ignited trees in forest fires and
magnetization in a ferromagnet close to a critical point,
the probability distribution of a global quantity 
(i.e. total energy dissipation, magnetization  and so forth) that has been
normalized to the first two moments follows a specific non Gaussian
 curve.  
This curve follows a form suggested by extremum statistics, 
which is specified by a single parameter $a$ ($a=1$ corresponds to the
Fisher-Tippett Type I (``Gumbel") distribution.)

Here, we present a framework for testing for extremal statistics in  
a global observable. In any given system, we wish to obtain $a$ in order
to distinguish between
the different Fisher Tippett asymptotes, and to compare with the above work. 
 The normalizations of the extremal curves are obtained as a function
of $a$. We find that for realistic ranges of data, 
  the various extremal distributions when normalized to the first two moments
  are difficult to distinguish. In addition, the convergence to the limiting
  extremal distributions for finite datasets is both slow and  
varies with the asymptote.
However, when the third moment is expressed as a function of $a$ this is found
to be a more sensitive method.
\end{abstract}

\section{Introduction}
The study of systems exhibiting non Gaussian
statistics is of considerable current interest (see e.g. \citet{sornette} and
references therein).
These statistics are often observed to arise in finite size  many body
systems exhibiting correlation over a broad range of scales; 
leading to emergent phenomenology such as  
self similarity and in some cases fractional dimension  \citep{redbook}.
The apparent ubiquitous nature of this behavior has led to interest in
self organized criticality \citep{bakbook,jensenbook} as a paradigm; 
other  highly correlated systems include those exhibiting fully developed
 turbulence.
 In solar terrestrial physics in particular, problems of interest include
 MHD turbulence in the solar wind and in the earth's magnetotail. Irregular or 
 bursty transport and energy release in the latter has recently led to
 complex system approaches such as SOC (see the review by Chapman and Watkins,
 Space. Sci. Rev., 2001).
 These complex systems are often characterized by a lack of scale,
and in particular, by the exponents of the power law probability
distributions (PDF) of patches of activity in the system. Examples of these
patches of activity 
are energy dissipated by avalanches in sandpiles,  vortices in turbulent fluids,
ignited trees in forest fires and magnetization in a ferromagnet close
to the critical point.
In the earth's magnetotail, patches of activity in the aurora as seen
by POLAR UVI have been used as a proxy for the energy released in bursty 
magnetotail transport in order to infer its scaling properties \citep{luiGRL,uritskyJGR}.
The challenge is to
distinguish the system from an  uncorrelated Gaussian process,
by demonstrating self similarity;
and to  determine the power law exponents. To do this directly is  nontrivial,
requiring  measurements of the individual patches or activity events
over many decades. Here we consider what may be a more
readily accessible measure, the statistics of a global average quantity such
as the total energy dissipation, magnetization  and so forth.

An important hypothesis that is the subject of this paper
is that the data  arise from an
extremum process; i.e. that some unknown selection process operates such that the observed
global quantity is dominated by the largest events selected from  ensembles of
individual `patches' of activity.
 This is a real possibility for two reasons. First,  
 measurements of physical
systems, and in particular, observations of natural systems,
inevitably incorporate instrumental thresholds and this may affect the statistics
of a global quantity comprising activity summed over patches. Second, there
has recently been considerable interest in a series of intriguing
results from turbulence experiments \citep{pinton1,pinton2,nature}, and numerical
models exhibiting  correlations (\citet{bramprl}, but see also
\citet{aji,comment,reply}). These studies reveal statistics of a global quantity (i.e. $E$)
that 
follows curves that are of the form of 
 one of the limiting extremal distributions:
 \citep{gumbook,ftippet}
   \begin{eqnarray} \nonumber
      P(E)=K (e^{y-e^y})^a\\\label{gum0}
	     y=b(E-s)
		    \end{eqnarray}
		    where $K,b$ and $s$ are obtained by normalizing to
		    the first two moments ($M_0=1$, $M_1=0$, $M_2=1$),
		    and
the single parameter $a$ appears to be close to the value $ \pi/2$. 

       For an infinitely large ensemble, there are
       two limiting distributions that we consider here. The Fisher-Tippett type I
       (or `Gumbel') extremal distribution
       is of the form (\ref{gum0}) but with $a=1$ and
       arises from selecting the largest events from 
ensembles with distributions that fall off exponentially or faster. 
Since we wish to construct a framework that could encompass
all highly correlated systems we also treat the
case where the distribution of `patches' is power law. An
example is the Potts model \citep{cardy} for magnetization where connected bonds form clusters,
the size of which is power law distributed at the critical point.
In this case the relevant extremal distribution is Fisher-Tippett
type II  (or 'Frechet').

Here we provide a framework for comparing data with Fisher Tippett type I and II
extremal curves. This essentially requires obtaining the normalizations of these
curves in terms of the moments of the data and ultimately as functions
of the single parameter $a$.

We find that the  curves of form (\ref{gum0})
which are obtained by normalizing to the
first two moments are
difficult to distinguish if $a$ is in the range $[1,2]$ or from Frechet 
curves given a realistic range of data.
Furthermore we demonstrate that slow convergence with respect to 
the size of the dataset, to the limiting
$a=1$ extremal distribution has the consequence that,
for a large but finite ensemble,
the extremal distribution of an uncorrelated Gaussian
process is indistinguishable from the $a=\pi/2$  curve.
To overcome these limitations we  suggest two much more sensitive
methods for determining 
whether or not the
curve is of the form (\ref{gum0}), and, if so, the corresponding 
value of $a$. These methods are based on  the third moment, and the peak
of the distribution, both of which we obtain here as a function of 
$a$.

\section{Extremum statistics: general results.}
To facilitate the work here we first develop some results from 
 extremum statistics (for further background reading see \citet{sornette,gumbook,bouchbook}). If the maximum $Q^*$ 
drawn from an ensemble of $M$ patches of activity $Q$
with distribution $N(Q)$ is
 $Q^*=max\{Q_1,..Q_M\}$,
then the probability distribution (PDF) for $Q^{*}$ is given by
\begin{equation}
P_m(Q^*)=MN(Q^*)(1-N_>(Q^*))^{M-1} \label{pmdef}
\end{equation}
where $M$ is the  number of patches in the ensemble and
\begin{equation}
N_>(Q^*)=\int_{Q^*}^\infty N(Q)dQ \label{nhat}
\end{equation}
We  now obtain $P_m$ for large $M,Q$. For general PDF $N(Q)$ we can write
(for appropriate choice of the function $g(Q^*)$):
\begin{equation}
(1-N_>)^{M}=e^{-Mg(Q^*)}
\end{equation}
and for small $N_>(Q^*)$ we have
\begin{equation}
g(Q^*)=-\ln(1-N_>(Q^*))\sim N_>+\frac{N_>^2}{2}\label{expN}
\end{equation}
We now  consider a characteristic value of $Q^*$,
namely $\tilde{Q}^*$, such that 
by definition
\begin{eqnarray}
Mg(\tilde{Q}^*)=q\label{xqdef}
\end{eqnarray}
so that
\begin{equation}
q=Mg(\tilde{Q}^*)\approx MN_>(\tilde{Q}^*)+M\frac{N_>^2(\tilde{Q}^{*})}{2}+\cdots\label{qdef}
\end{equation}
We now expand $g(Q^*)$ about $\tilde{Q}^*$ to obtain 
\begin{eqnarray}
g(Q^*)=g(\tilde{Q}^*)+g'(\tilde{Q}^*)\Delta Q^* +
\frac{g''(\tilde{Q}^*)}{2}(\Delta Q^*)^2
+\cdots\label{taylorg}
\end{eqnarray}
and from (\ref{expN}) we have
\begin{eqnarray}
g'(Q^*)=-N(Q^*)-N(Q^*)N_>+\cdots\\
g''(Q^*)=-N'(Q^*)-N'(Q^*)N_>+N^2(Q^*)+\cdots
\end{eqnarray}
where $g',g''$ denote differentiation with respect to $Q^*$,  $\Delta Q^*=Q^*-\tilde{Q}^*$, and we have used 
$N'_>=dN_>/dQ^*=-N$.
Inverting expansion (\ref{qdef}) gives
\begin{equation}
MN_>(\tilde{Q}^*)=
q\left[1-\frac{q}{2M}+\cdots\right]\approx M\left(1-e^{-\frac{q}{M}}\right)\label{nn}
\end{equation}
We obtain from (\ref{expN}) and its derivatives with respect to
$Q^*$:
\begin{eqnarray}\nonumber
g(\tilde{Q}^*)=\frac{q}{M}\left(1-\frac{q}{2M}\right)+\frac{1}{2}\left(\frac{q}{M}\right)^2+\cdots\\
=\frac{q}{M}+0\left(\frac{q}{M}
\right)^3
\end{eqnarray}
which to relevant order is consistent with (\ref{xqdef}), and
\begin{eqnarray}
g'(\tilde{Q}^*)=-N(\tilde{Q}^*)\left[1+\frac{q}{M}+\cdots\right]
\end{eqnarray}
For $q$ finite as $M\rightarrow \infty$ this gives $g'(\tilde{Q}^*)=-N(\tilde{Q}^*)$
and $MN_>(\tilde{Q}^*)=q$.

We can now consider the extremal statistics of specific  PDF $N(Q)$,
and importantly show that $P_m(Q^*)$ can be written in the universal form (\ref{gum0}).
\subsection{Gaussian and Exponential $N(Q)$}
If $N(Q)$
falls off sufficiently fast in $Q$, i.e. is
Gaussian or exponential it is sufficient to
consider lowest order only in (\ref{expN}) giving  $g(Q^*)\sim N_>$ \citep{gumbook,bouch}
 and $q=MN_>(\tilde{Q}^*)$. Expanding (\ref{nhat}) in $Q^*$ near $\tilde{Q}^*$ gives
to this order:
\begin{eqnarray}\nonumber
MN_>(Q^*)=M\int^\infty_{\tilde{Q}^*}
N(Q)dQ-MN(\tilde{Q}^*)\Delta Q^*\\\label{extra1}
=q\left[1- \frac{MN(\tilde{Q}^*)}{q} \Delta Q^* + \cdots \right]
\approx q e^{-M\frac{N(\tilde{Q}^*)}{q}\Delta Q^*}
\end{eqnarray}
Expanding $N(Q)$ about $Q^*$ yields

\begin{eqnarray}\nonumber
N(Q^*)=N(\tilde{Q}^*)\left[1+\frac{N'(\tilde{Q}^*)}{N(\tilde{Q}^*)}
\Delta Q^* +\cdots\right]\\\label{extra2}
\approx N(\tilde{Q}^*)e^{\frac{N'(\tilde{Q}^*)}{N(\tilde{Q}^*)}
\Delta Q^*}
\end{eqnarray}
As to this order $(1-N_>)^{M-1} \approx e^{-MN_>}$
we then have from (\ref{pmdef})
\begin{eqnarray}\nonumber
 P_m(Q^*) = MN(Q^*)(1-N_>(Q^*))^{M-1}\approx MN(Q^*) e^{-MN_>}\\
\sim (e^{u-e^u})^a\label{gumgum}
\end{eqnarray}
with
\begin{eqnarray}
a=-\frac{N'(\tilde{Q}^*)N_>(\tilde{Q}^*)}{N^2(\tilde{Q}^*)} \label{agumb}
\end{eqnarray}
and
\begin{eqnarray}
u=\ln \left(\frac{M N_>(\tilde{Q^*})}{a} \right)
-\frac{N(\tilde{Q}^*)}{N_>(\tilde{Q}^*)}\Delta Q^*
\label{expu}
\end{eqnarray}

Since throughout we are considering $\tilde{Q}^*$ large ($M\rightarrow \infty ,q$ finite)
we have the effective value of $a$ as that given by (\ref{agumb}) in the limit
$\tilde{Q}^*\rightarrow\infty$.
For $N(Q)$ exponential the above gives $a=1$. In the particular case of the
exponential all the summations
which in the above we have truncated can be resummed exactly and give $a\equiv 1$,
recovering the result of \citet{bouch}.

For $N(Q)$ Gaussian we cannot obtain $a$ exactly in this way
but as we shall see it is instructive
to make an estimate. Given $N(Q)=N_0 \exp(-\lambda Q^2)$
and expanding equations (\ref{extra1}),  (\ref{extra2}) and (\ref{gumgum}) 
to next order we obtain
\begin{eqnarray}\nonumber
P_m=\bar{P_m}e^{R(u)}\\
R=-\frac{\ln^2(q)}{4\lambda \tilde{Q}^{*2}}\label{Reqn}
+\bar{u}\left(1+\frac{2 \ln (q)}{4\lambda \tilde{Q}^{*2}}\right)
-\frac{\bar{u}^2}{4\lambda \tilde{Q}^{*2}}
-e^{\bar{u}}
\end{eqnarray}
where we have used $u=-2\lambda \tilde{Q}^*\Delta Q^*$ and $\bar{u}=u+\ln(q)$.
To lowest order in $\Delta Q^*/\tilde{Q}^*$ (i.e. $\tilde{Q}^*\rightarrow \infty$)
we have a universal PDF with $a=1$, but to next order, that is, neglecting only
the term in $\bar{u}^2$  in (\ref{Reqn}) we have a universal distribution of form (\ref{gum0},\ref{gumgum}) with
\begin{equation}
a\equiv \left(1+\frac{2 \ln (q)}{4\lambda \tilde{Q}^{*2}}\right)\neq 1
\end{equation}

\subsection{Power law $N(Q)$}

The PDF of patches $N(Q)$ may however be a power law
and in this case it will fall off sufficiently slowly with $Q$ that we need
to go to next order as in
(\ref{qdef}). If we consider a normalizable source PDF
\begin{equation}
N(Q)=\frac{N_0}{(1+Q^2)^k}\label{power1}
\end{equation}
then for large $Q$ ($Q>>1$) we have $N(Q) \sim N_0/Q^{2k}$ and then
using (\ref{nhat}) and (\ref{qdef})
\begin{equation}
\tilde{Q}^*N(\tilde{Q}^*)=
(2k-1)N_>(\tilde{Q}^*)=(2k-1)\frac{q}{M}(1-\frac{q}{2M})
\label{mystery}
\end{equation}
which with the above general expressions for  $g(\tilde{Q}^*)$
and its derivatives  
substituted into (\ref{taylorg}) gives an expression for $g(Q^*)$
\begin{equation}
g(Q^*)=\frac{q}{M}\left[1-(2k-1)\frac{\Delta Q^*}{\tilde{Q}^*}+
k(2k-1)(\frac{\Delta Q^*}{\tilde{Q}^*})^2\cdots\right]\label{powerg}
\end{equation}
We also require an expression for $N(Q^*)$, again expanding about
$\tilde{Q}^*$ and obtaining the derivatives of $N(\tilde{Q}^*)$ from those
of $g(\tilde{Q}^*)$ and via (\ref{nn}) gives
\begin{equation}
N(Q^*)=N(\tilde{Q}^*)\left[1-2k\frac{\Delta Q^*}{\tilde{Q}^*}+k(2k+1)
(\frac{\Delta Q^*}{\tilde{Q}^*})^2\right]
\end{equation}
which can be rearranged as
\begin{equation}
N(Q^*)=N(\tilde{Q}^*)e^{\left[-2k\frac{\Delta Q^*}{\tilde{Q}^*}
+k(\frac{\Delta Q^*}{\tilde{Q}^*})^2\right]}
\end{equation}
After some algebra (\ref{powerg}) can be rearranged to give
\begin{equation}
Mg(Q^*)=qe^{\left[-(2k-1)\frac{\Delta Q^*}{\tilde{Q}^*}
+\frac{2k-1}{2}(\frac{\Delta Q^*}{\tilde{Q}^*})^2\right]}
\end{equation}
These two expressions combine to finally give
\begin{equation}
P_m(Q)\equiv P_m(Q^*)\sim (e^{\bar{u}-e^{\bar{u}}})^a
\end{equation}
with
\begin{equation}
\bar{u}=-\ln(a)-\ln(q)-(2k-1)\frac{\Delta Q^*}
{\tilde{Q}^*}(1-\frac{\Delta Q^*}{2\tilde{Q}^*})\label{baru}
\end{equation}
and
\begin{equation}
a=\frac{2k}{2k-1}\label{afork}
\end{equation}
To lowest order, neglecting the $(\Delta Q^* / \tilde{Q}^*)^2$
term 
 (\ref{baru}) reduces to 
(\ref{expu}).

 Hence  a power law PDF
has maximal statistics
$P_m(Q)$ which, when evaluated to next order,  can be written in the form of a universal
curve (i.e. of form (\ref{gum0},\ref{gumgum})) with a correction that is non negligible
  at the asymptotes. This can be seen \citep{jenkinson,bouchbook} to be consistent with
  the well known result due to Frechet where (following the notation of \citet{bouchbook})
   if we have PDF 
\begin{equation}
N(x)\sim \frac{1}{\mid x\mid^{1+\mu}}
\end{equation}
then
\begin{eqnarray}
N_>\sim \frac{1}{x^\mu}\\\nonumber
P_m(x^*)=\frac{\mu}{(x^*)^{1+\mu}} e^{-\frac{1}{(x^*)^\mu}}
\end{eqnarray}
which we can write in the form
\begin{eqnarray}
P_m(x^*)=\mu e^{\frac{\mu+1}{\mu}\ln(\frac{\mu+1}{\mu})}(e^{u -e^u})^a\\
u=-\mu \ln(x^*)-\ln\left(\frac{\mu+1}{\mu}\right)\label{mukdef}
\end{eqnarray}
which is of universal form (\ref{gum0},\ref{gumgum}) in $u$.
Noting that here
$\mu=2k-1$ and $a=(\mu+1)/\mu$ and that to second order
\begin{equation}
\frac{\Delta Q^*}{\tilde{Q}^*}(1-\frac{\Delta Q^*}{2
\tilde{Q}^*})=\ln\left(1+\frac{\Delta Q^*}{\tilde{Q}^*}\right)\label{lnQ}
\end{equation}
we simply identify $1+\Delta Q^*/\tilde{Q}^*$ with $\tilde{x}^*$
to obtain (\ref{baru}).
To next order in $\Delta Q^*/\tilde{Q}^*$ the analogue of (\ref{baru}) still
yields the right hand side of (\ref{lnQ}).
\subsection{Convergence to the limiting distributions}

The above results should be contrasted with the derivation of Fisher and Tippett
\citep{ftippet}.
Central to  \citep{ftippet} and later derivations is that 
a single ensemble of $NM$ patches has the same statistics as
the 
 $N$ ensembles 
(of $M$ patches), of which it is comprised.  
The fixed point of the resulting functional equation \citep{barrow} for arbitrarily large $N$ and $M$ 
is $a=1$ for the exponential
and Gaussian PDF, and the Frechet result for power law PDF.
Here, we consider a finite sized system so that although the number
of realizable ensembles of the system can be taken arbitrarily large, the
number of patches $M$ per ensemble is always large but finite.
Importantly, the rate of convergence with $M$ depends on the
PDF $N(Q)$. For an exponential or power law PDF
we are able to resum the above expansion exactly to obtain
$a$; and convergence will then just depend on terms $O(1/M)$ and above.
This procedure is not possible for $N(Q)$ Gaussian, instead we 
consider the characteristic $Q^*$, that is $\tilde{Q}^*$ 
which for  $M$ arbitrarily large should be large also.
Rearranging
(\ref{qdef}) to lowest order for  $N(Q)=N_0 \exp(-\lambda Q^2)$
yields $\sqrt \lambda \tilde{Q}^* \sim
\sqrt{\ln(M)}$ implying significantly slower convergence.
This is further discussed in \citet{sornette} (pp. 19-21).

The
extremal distributions are  thus essentially a family
of curves that are approximately
of universal form (\ref{gum0},\ref{gumgum}) and are asymmetric with a handedness
 that just depends on the sign of $Q$; we have assumed $Q$ positive
 whereas one could choose $Q$ negative in which case
 $N(Q)\rightarrow N(\mid Q \mid)$. This would correspond to, say, power absorbed, rather
than emitted, from a system.
The  single parameter $a$ that distinguishes
the extremal PDF then just depends on the  PDF of
the individual events. For $N(Q)$ exponential we then recover exactly the
well known result  \citep{gumbook,bouch} $a=1$.  
For a power law PDF $a$ is determined by $k$ via (\ref{afork}).
We have also demonstrated that for a Gaussian PDF with finite but large $M$ and
$N$ that $a \neq 1$ 
and will explore the significance of this in Section 3.1.

\section{Normalization to the first two moments}

To compare these curves with data we need
$P(\bar{Q})\equiv P_m(Q^*)$ in normalized form. 
This has moments
\begin{equation}
M_n=\int_{-\infty}^\infty y^n \bar{P}(y) dy\label{moment}
\end{equation}
which we will obtain as a function of $a$ and then insist that $M_0=1$,  $M_1=0$ and $M_2=1$.

 Setting  $M_1=0$  (and $M_0=1$, $M_2=1$) in our analysis
 of extremal distributions does not require any assumptions
 about the form of the PDF except that the moments exist.  
 It will allow us to write the analytically obtained extremal 
 distributions as functions
 of single parameter $a$.

\subsection{Extremal distributions arising from Gaussian and exponential $N(Q)$}

For Gaussian and exponential  PDF we have
\begin{eqnarray}
\bar{P}(y)=K  (e^{u-e^u})^a\label{gumy}\\
u=b(y-s)\label{expu2}
\end{eqnarray}
This has moments
which converge for all $n$. 
From Appendix A we have that the $n^{th}$ moment:
\begin{equation}
M_n=\frac{1}{b}\int^\infty_{-\infty} 
\bar{P}(y) d\eta \frac{[\ln(a)+bs-\eta]^n}{b^n}
=Ke^{-a\ln(a)}\frac{d^n}{da^n}\Gamma(a)
\end{equation}
where $\eta=\ln(a)-u$.

To normalize we insist that $M_0=1$,  $M_1=0$ and $M_2=1$.
The necessary integrals can be expressed in terms of derivatives
of the Gamma function $\Gamma(a)$ (\citet{gamref})
and
we obtain in Appendix A:

\begin{eqnarray}\nonumber
b^2= \Psi'(a)\nonumber\\
K=\frac{b}{\Gamma(a)}e^{a \ln(a)}\label{gumnorm}\\
s= - \frac{(\Psi(a)-\ln(a))}{b}\nonumber
\end{eqnarray}
where
\begin{eqnarray}\nonumber
\Psi(a)=\frac{1}{\Gamma(a)}\frac{d\Gamma(a)}{da}\\\nonumber
\Psi'(a)=\frac{d\Psi}{da}
\end{eqnarray}

The ambiguity in the sign of $b$ (and hence $s$) corresponds to
the two solutions for $P(\bar{Q})$ for positive and negative $Q$.

We can now plot the  curves, that is, normalized to the
first two moments and these are shown in Figure 1. Experimental measurements of
a global PDF $P(E)$ normalized to $M_0$
would be plotted $M_2 P$ versus $(E-M_1)/M_2$.
 In the main plot we show normalized  distributions of the form (\ref{gum0},\ref{gumgum})
  for $a=1,\pi/2$ and
  $2$. It is immediately apparent that the
  curves are difficult to distinguish over several decades
  in $\bar{P}(y)$ and thus to obtain a good
  estimate for $a$, the numerical or real experiments would require
  good statistics over a dynamic range of about 4 decades, something which is not
  readily achievable.

\begin{figure}
  \includegraphics[width=15cm]{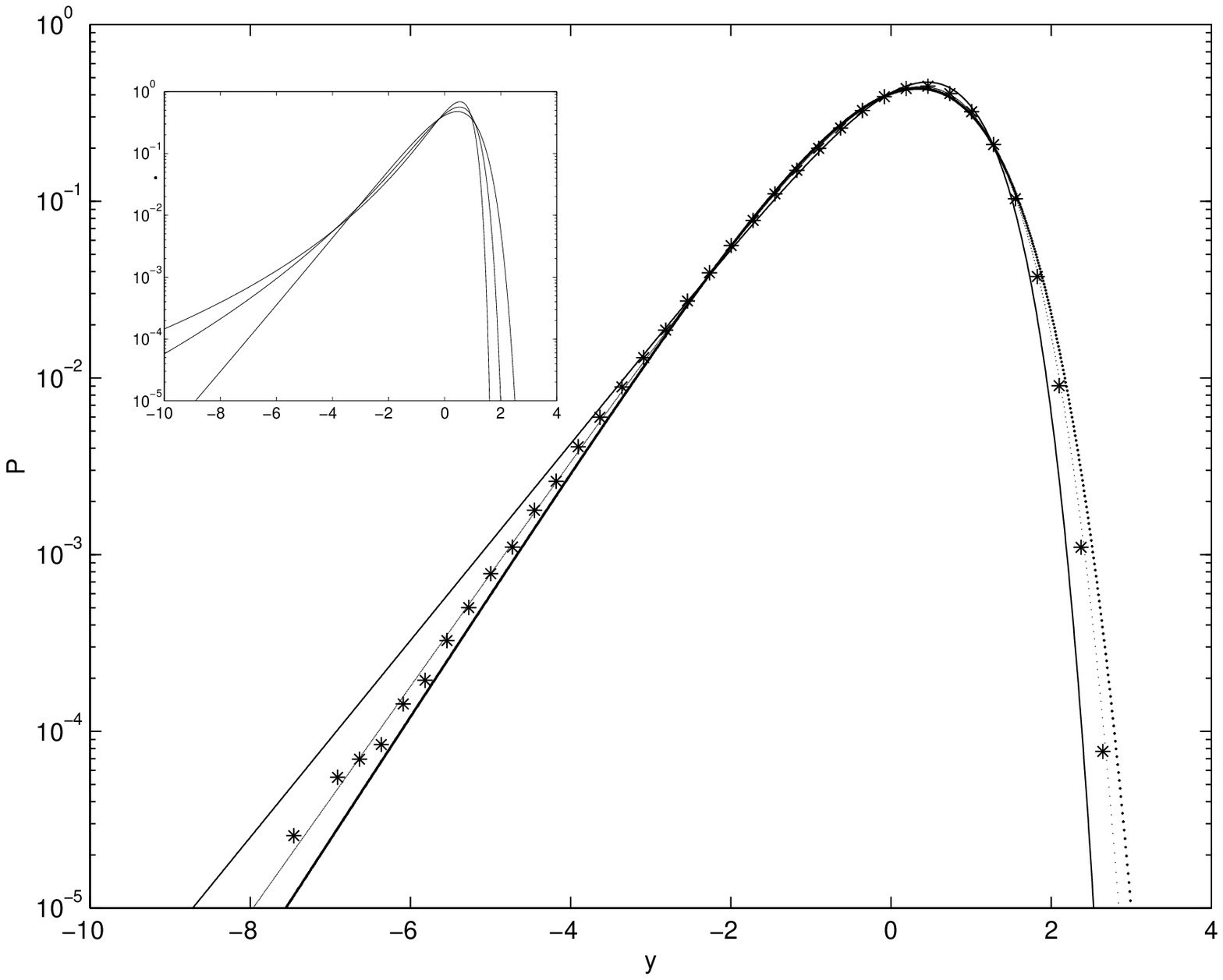}
\caption{Curves of the form (\ref{gum0}) for $a=1,\pi/2,2$.
Overlaid (*) are the numerically
 calculated extremal statistics of an uncorrelated Gaussian process (see text), and inset
    for comparison are Frechet curves plotted on the same scale (see Fig. 2).
    }

    \end{figure}

On Figure 1 we have also over plotted (*) the extremal PDF of ensembles of 
 uncorrelated numbers that are Gaussian distributed, calculated
 numerically. We randomly select $M$ uncorrelated variables
 $Q_{j},j=1,M$ and to specify the handedness of the extremum distribution,
 the $Q_j$ are defined negative and $N(\mid Q \mid)$
 is normally distributed. This would physically
  correspond to a system where the global quantity $\bar{Q}$ is negative,
   i.e. power consumption in a turbulent fluid, as opposed to power generation.
   To construct the global PDF we generate  $T$ ensembles, that is select
 $T$ samples of the largest negative number $Q^*_i=min\{Q_1..Q_M\}, i=1,T$.
 For the data shown in the figure $M=10^5$ and $T=10^6$; this gives
 $\sqrt \lambda \tilde{Q}^* \sim \sqrt{\ln(M)}\simeq 3$ so that for the
 Gaussian we are far from the $a=1$ limit \citep{ftippet}.
The numerically calculated PDF lies close to $a=\pi/2$.
 Such a value of $a$ on these  curves thus does not give direct
  evidence of a {\em correlated} process; in addition it is necessary to
establish that the data considered do not arise as the result of an extremal process.

Generally, plotting data in this way is an 
insensitive method for determining $a$ 
and thus distinguishing the statistics of the underlying physical process.
The question of interest is whether we can determine the form of the
curve, and the value of $a$ from data with a reasonable dynamic range; we address this
question in
section IV.

\subsection{Frechet distributions arising from power law $N(Q)$}

For power law  PDF (\ref{power1}) we use
the Frechet distribution which we first write as:
\begin{eqnarray}
P(Q^*)=K(e^{u-e^u})^a\\
u=\alpha+\beta \ln(1+\frac{Q^*}{\tilde{Q}^*})\label{frechu}
\end{eqnarray}
which reduces to the form of (\ref{expu2}) for 
$\Delta Q^*/\tilde{Q}^* \ll 1$. 
From (\ref{baru}), (\ref{power1}) and (\ref{mukdef}) we identify
\begin{equation}
\beta=-\mu=-(2k-1)
\end{equation}
The procedure of normalizing to the moments
is only valid provided that they exist.
For the power law PDF (\ref{power1}) we have (see also \citet{bury}):
\begin{eqnarray*}
M_n=\int_0^\infty \frac{Q^n H(Q)dQ}{(1+Q^2)^k}
\end{eqnarray*}
Which converges for $Q\rightarrow 0$ and for $Q\rightarrow \infty$
\begin{eqnarray*}
M_n\sim\int^\infty \frac{Q^n H(Q)dQ}{Q^{2k}}
\end{eqnarray*}
which  if $H(Q)\rightarrow H_0$ as $Q\rightarrow \infty$
\begin{eqnarray*}
M_n \rightarrow \int^\infty
\frac{dQ}{Q^{2k-n}}\simeq \frac{1}{Q^{2k-n-1}}\mid_{Q\rightarrow\infty}
\end{eqnarray*}
which converges if $ 2k>n+1$.

We now evaluate the moments.
Again we insist that $M_0=1$,  $M_1=0$ and $M_2=1$ and in Appendix
B obtain:
\begin{eqnarray}\nonumber
\alpha=-\beta \ln\left(\frac{a^{\frac{1}{\beta}}}{\Gamma(1+1/\beta)}\right)\\
K=\pm \beta a^a \left[\Gamma(1+\frac{2}{\beta})-\Gamma^2(1+\frac{1}{\beta})
\right]^\frac{1}{2}\label{frechnorm}\\
\tilde{Q^*}=\frac{\Gamma(1+\frac{1}{\beta})}{\left[\Gamma(1+\frac{2}{\beta})-\Gamma^2(1+\frac{1}{\beta})
\right]^\frac{1}{2}}\nonumber
\end{eqnarray}
where $\beta = - (2k-1)$.
The normalization constants are thus also expressible as functions of 
$a=2k/(2k-1)$.

For convergence, these curves
exist for power law of index $\infty > 2k > 3$ i.e.
$ 1 < a < 3/2$. 
This is
 significant since processes exhibiting intermittency as a consequence
of long range correlations typically have $k$ lower than this \citep{jensenbook},
and we will consider alternative methods in  section 5.

In Figure 2 we plot the normalized Fisher Tippett type II or
Frechet PDF for $k=2,5,100$  
and for comparison the Fisher Tippett type I ('Gumbel')
PDF with $a=1$. From (\ref{afork}) $a=1$
corresponds to $k\rightarrow\infty$ and it is straightforward to
demonstrate from the algebra that in this limit, the normalized
Frechet PDF tends to Gumbel's asymptote $a=1$.
Hence on this plot we see that for $k=100$ these are indistinguishable
and differences between the Frechet and Gumbel PDF
only  appear on such a plot around the mean for $k<3$ approximately. This demonstrates that
these extremal curves arising from an uncorrelated Gaussian, exponential or power
law $N(Q)$ will all be difficult to distinguish from the  curve (\ref{gum0},\ref{gumgum})
with $a\neq 1$. We now consider more sensitive methods to determine $a$.

\begin{figure}
  \includegraphics[width=15cm]{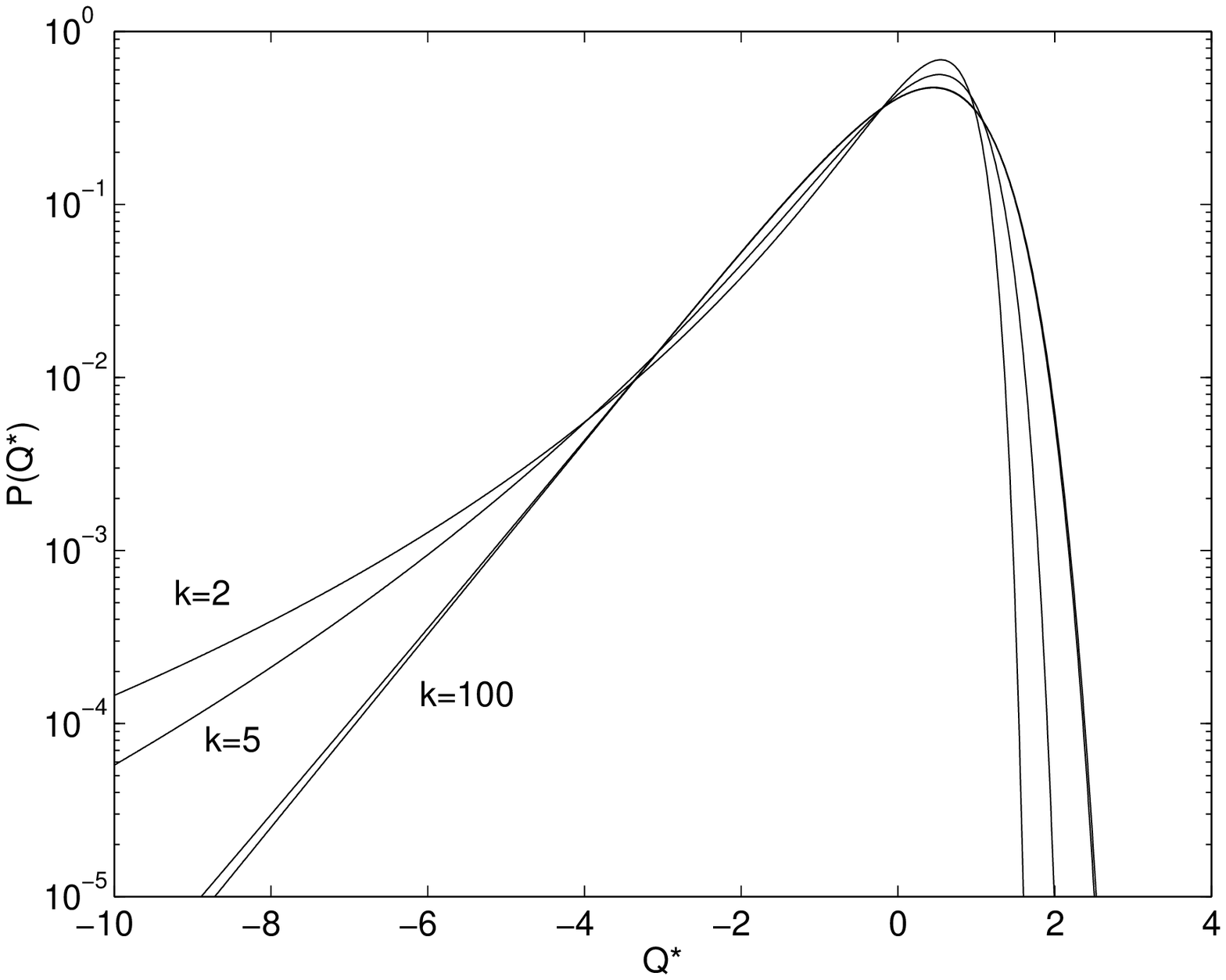}
  \caption{Frechet PDF normalized to the first two moments for
  PDF $N(Q)=1/(1+Q^2)^k$,  $k=2,5,100$.
  }

\end{figure}

\section{Sensitive indicators of $a$; the mean and the third moment}
The question  of interest is whether we can determine 
$a$ with sufficient accuracy from data with
 a reasonable dynamic range. 
 We consider two possibilities here.

First, a uniformly sampled process will have the most statistically significant
values on the extremal curve near the peak, and in particular,
from the figures we see that 
the Frechet distributions for small $k$ will be most easily
distinguished in this way. 
For the Frechet PDF the peak is at $u=0$, that is, it has coordinates 
\begin{eqnarray}
\bar{P}_m=\frac{K}{e^a} \;\;\;\;
\bar{y}=\tilde{Q}\left[e^{-\frac{\alpha}{\beta}}-1\right]
\end{eqnarray}
on the normalized curve 
with $K, \tilde{Q}, \alpha, \beta$  known as functions of $a$ from Appendix B.
The coordinates of the peak of the
PDF from the data plotted with $M_0=1, M_1=0$
and $M_2=1$ can thus be graphically inverted to give an estimate of $a$. 

For  PDF that are power law with large $k$,  exponential 
or Gaussian, we consider 
the normalized extremal PDF; then the coordinates of the maximum
of $\bar{P}(y)$ is at $u=0$, $y=s$, that is: 
\begin{equation}
\bar{P}_m=\frac{K}{e^a}=\frac{\sqrt{\Psi'(a)}e^{-a(1-\ln(a))}}{\Gamma(a)}
\end{equation}
with $K,s$ from (\ref{gumnorm}).
These can again be graphically inverted to obtain $a$; Figure 3 shows
$\bar{P}$ and $\bar{y}$ versus 
  $k$ for the Frechet PDF.

\begin{figure}
  \includegraphics[width=10cm]{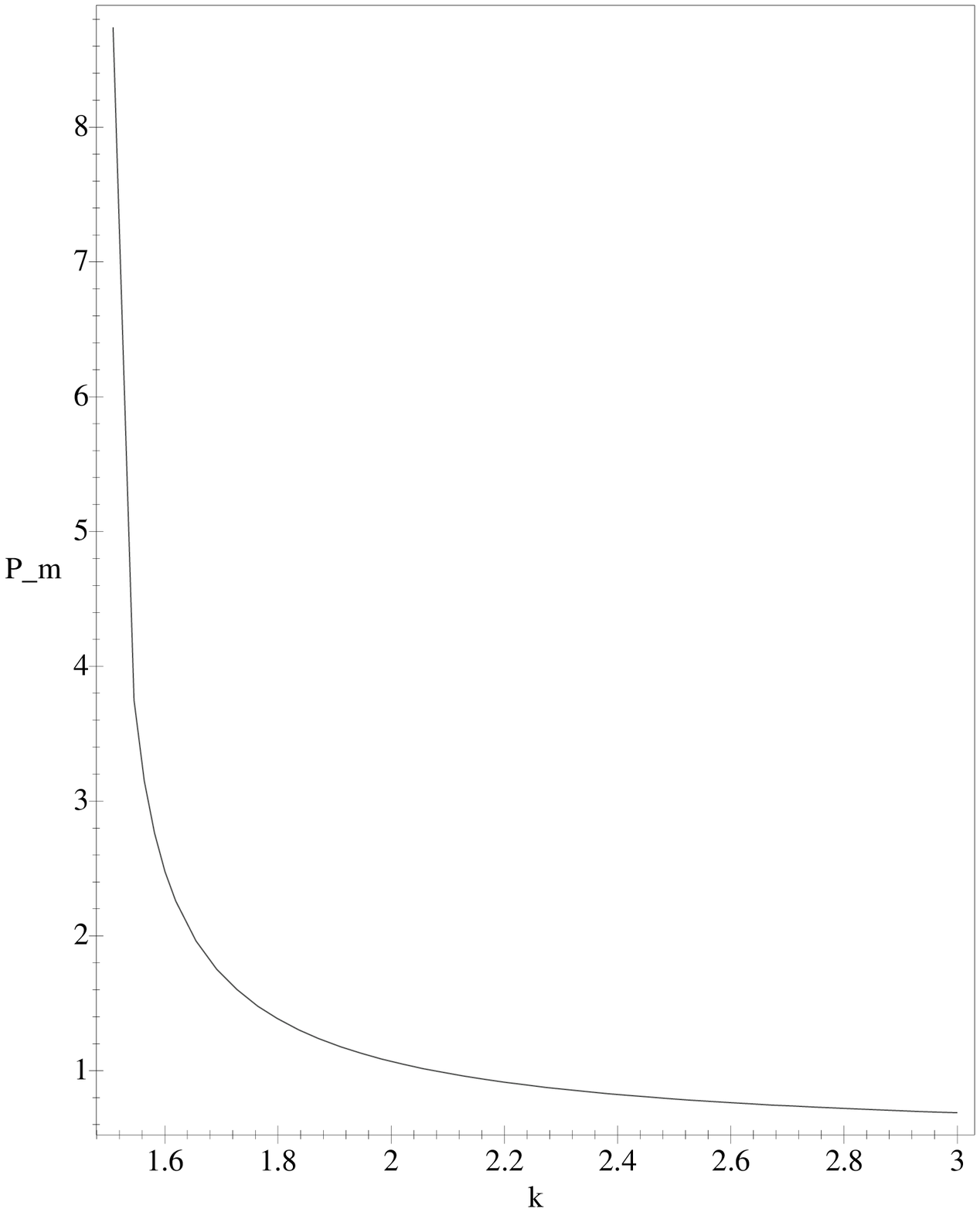}
  \includegraphics[width=10cm]{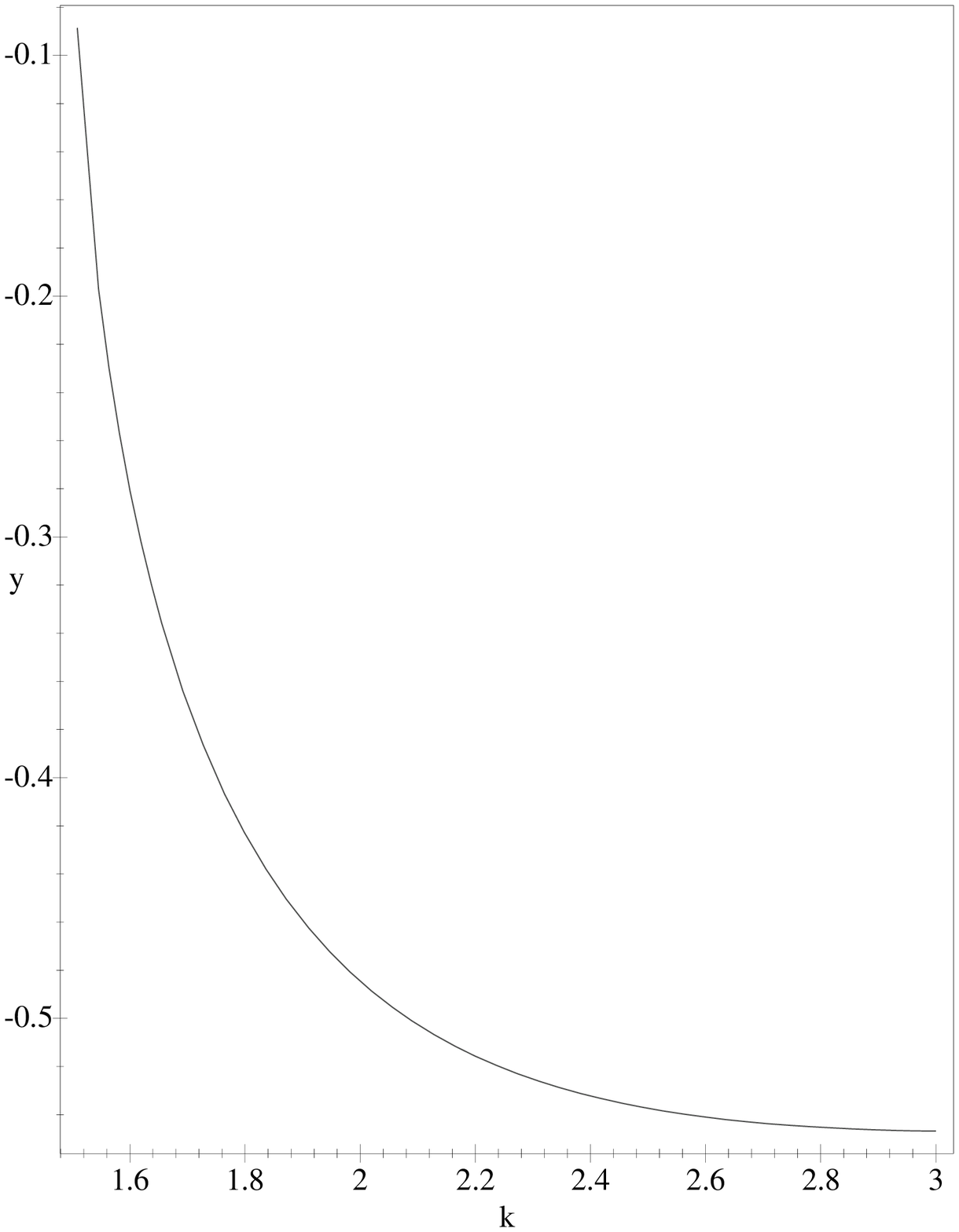}
  \caption{The peak (a) and its location (b) as a function of $k$ for Frechet curves.}

\end{figure}

A more sensitive indicator
may be the third moment of $\bar{P}$  of the
 curve  (\ref{gum0},\ref{gumgum}) which after some algebra
(Appendix A)
can be written as
\begin{equation}
M_3=-\frac{\Psi''(a)}{(\Psi'(a))^\frac{3}{2}}
\end{equation}
for a Gaussian or exponential PDF i.e. with (\ref{expu2})
and
\begin{equation}
M_3=\frac{\left[\Gamma(1+\frac{3}{\beta})-3\Gamma(1+\frac{2}{\beta})
\Gamma(1+\frac{1}{\beta})+2\Gamma^3(1+\frac{1}{\beta})\right]}
{\left[\Gamma(1+\frac{2}{\beta})-\Gamma^2(1+\frac{1}{\beta})\right]^\frac{3}{2}}
\end{equation}
for a power law  PDF (Appendix B) i.e. with (\ref{frechu}); the latter then converging
for $k>2$.
Again these refer to one of the two possible solutions for $P(\bar{Q})$;
the other solution 
 corresponding to $y\rightarrow -y$ ($Q^*\rightarrow -Q^*$)
in equations (\ref{expu2},\ref{frechu}) which in turn gives 
$M_3 \rightarrow -M_3$. 

The third moment is plotted versus $a$ and $k$ respectively
in Figure 4 for the Gumbel and Frechet curves.
Inspection of Figure 4  shows that over 
most of the range, $M_3$ is more sensitive than $\bar{P}$.
For Frechet curves, 
$M_3$ only has convergence for relatively large $k$
($k>2, a<4/3$); for smaller $k$, $\bar{P}$ can distinguish 
the Frechet distributions ($k>3/2,a <3/2$ for convergence).

\begin{figure}
  \includegraphics[width=10cm]{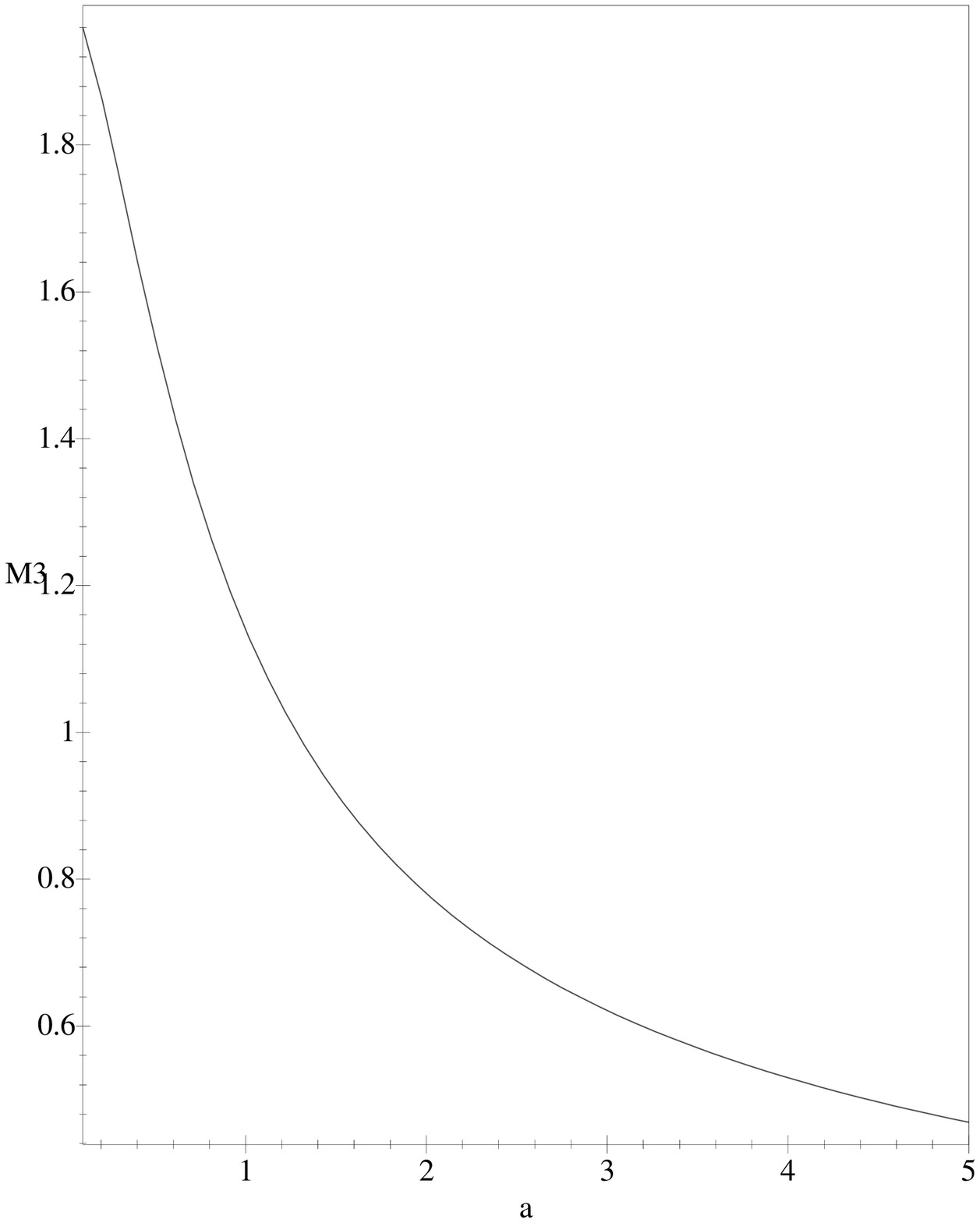}
    \includegraphics[width=10cm]{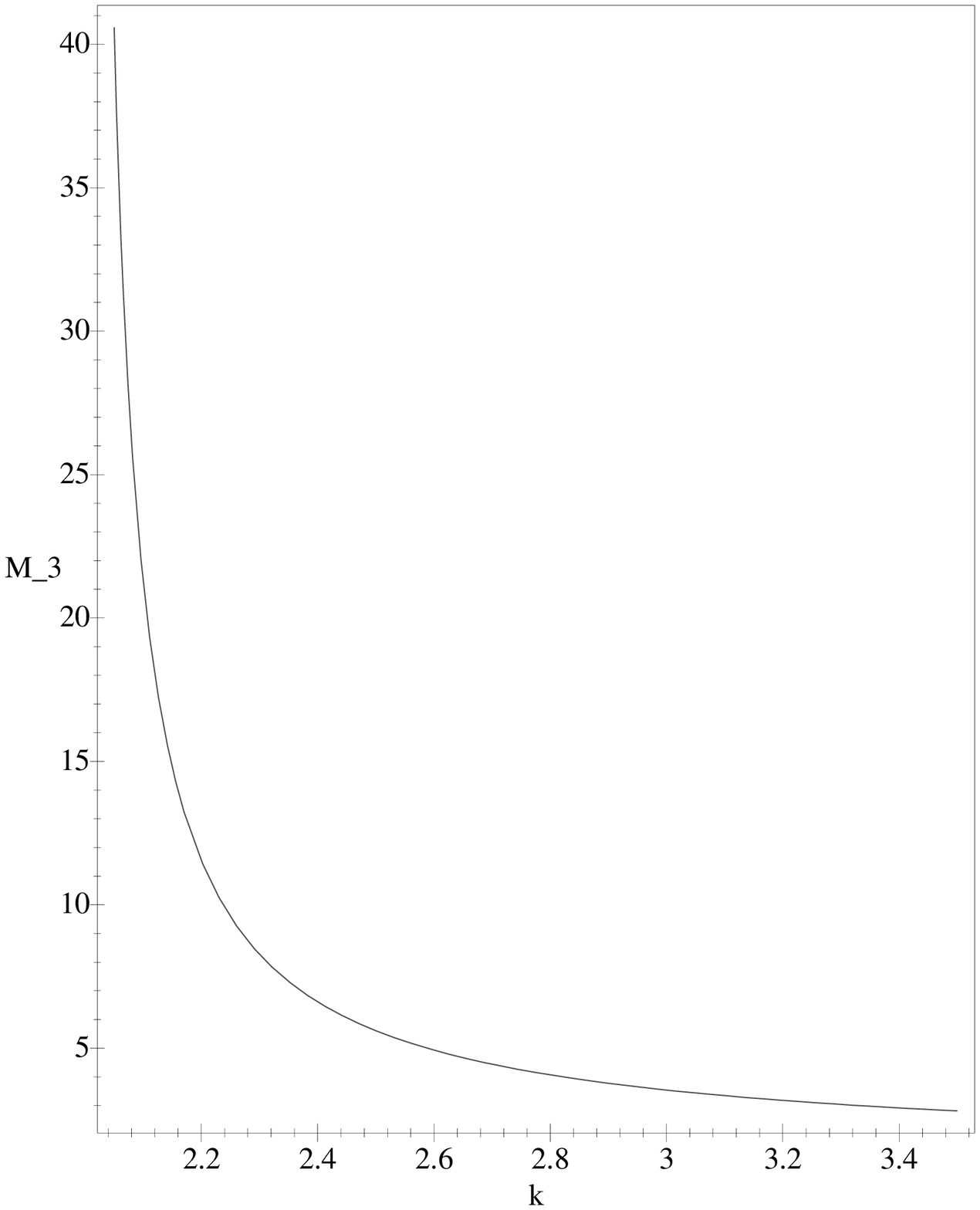}
\caption{The third moment as a function of $a$ for (a) curves of form (\ref{gum0}) and (b)
Frechet curves.}

      \end{figure}

\section{A method for small $k$}

For $N(Q)$ power law, we can only use the properties
of the normalized Frechet
PDF above for $k>3/2$. If $k$ is smaller than this the second moment
will not exist.
We can however obtain a useful result
for $k>1$ by using the first moment only, i.e. by insisting $M_0=1, M_1=0$.
We need another condition and can arbitrarily insist $P(u=0)=1$
(insisting that all the maxima of the Frechet PDF have the same height)
which  gives the condition

\begin{equation}
Ke^{-a}=1
\end{equation}

From  B6 and B5
\begin{equation}
\frac{K \tilde{Q}^{*}}{\beta g^{1/\beta} a^a}=1
\end{equation}
which, with $g^{1/\beta}=\Gamma (1 + 1/\beta)$ from Appendix B gives
$\tilde{Q}^{*}$ in terms of $a$ and $\beta$ (or $k$). 
Similarly we use (B5); $g = a e^{\alpha}$ to obtain $\alpha$ in terms of
$a$ and $\beta$.

This then gives

\begin{eqnarray*}
P_m(Q^{*})=K\left(e^{u-e^{u}}\right)^a\\
u=\alpha+\beta \ln(1+\frac{\Delta Q^{*}}{\tilde{Q}^{*}})\\
\alpha=\beta \ln\left(\Gamma(1+\frac{1}{\beta})\right)
-\ln(a)\\
\tilde{Q}^{*}=\beta e^{a(\ln(a)-1)}\\
K=e^a
\end{eqnarray*}

\section{Conclusions}

Recent work has suggested that the probability distribution
of some global quantity, such as total power needed to drive rotors
at constant velocity in a turbulent fluid, or total
magnetization in a ferromagnet slightly off the critical point, when normalized to
the first two moments, follows a non-Gaussian, universal curve. This curve is of the same
form as that found from the extremal statistics of a process that falls off exponentially
or faster at large values (i.e. Fisher-Tippett type I or `Gumbel'); but whereas for an extremal process the parameter specifying
the curve $a=1$, for the  correlated processes $a>1$. 

In this paper a framework has
been developed to compare data with Fisher-Tippett type I (`Gumbel') and type
II (`Frechet') asymptotes by obtaining the curves, and their normalizations,
as a function of a single parameter $a$. We find:

\begin{enumerate}

\item The Fisher Tippett type I and type II curves and their corresponding
values of $a$ are most easily distinguished by considering either the third moment,
or the position of the peak, as functions of $a$, the functional forms for which 
are given here.

For realistic ranges of data, simply comparing curved normalized to the first two moments as for example in
\citep{nature,bramprl}
is insufficient to adequately distinguish either curves of the form
of type I ('Gumbel')  but
with $a$ values in the range $[1,2]$,  or most type II ('Frechet') curves.

\item Convergence to the limiting form of the extremal curve $a=1$ (Gumbel's asymptote \citet{ftippet})
is sufficiently slow for an uncorrelated Gaussian that for a large but realistic size
of dataset one obtains $a\approx\pi/2$. Data which falls on this curve
is thus not sufficient to unambiguously distinguish a global
observable of a system that
 has correlations
\citep{nature,bramprl}, from that of an uncorrelated, extremal process.

\end{enumerate}

Comparison with data is then facilitated in the following way. First, the data
distribution is normalized to $M_0$ (to obtain the PDF $N(Q)$ say).
Second, the data is plotted on semilog axes under the following
normalization: $N(Q)\times M_2$ versus  $(Q-M_1)/M_2$. Any Gaussian 
PDF on such a plot will fall on a single inverted
parabola; similarly any Gumbel (Fisher Tipett I) process
will fall on a single curve. Finally, $M_3$
is calculated for the data; we then can compare the data with an extremal process 
by inverting $M_3(a)$ obtained here for a Fisher Tipett type I or II        
distribution. Overlaying these curves (augmented by other
quantitative comparisons) then essentially constitutes a fitting
procediure; but importantly, in addition the value of $a$ 
is related to the underlying distribution as we have discussed.

This and related techniques will have relevance in particular for regions
where transport is dominated by turbulence, in the solar wind and magnetosphere
in circumstances where multipoint and long time interval in situ measurements
are difficult to obtain.

\begin{acknowledgements}

The authors would like to thank G. King, M. P. Freeman,  D. Sornette  and J. D. Barrow
for illuminating discussions.
SCC was supported by PPARC.\\
\end{acknowledgements}

\appendix
\section{Moments of the Gumbel
distribution and the normalization $b$, $K$ and $s$ as a function of $a$.}

We consider a family of curves of the form 
\begin{equation}
P(y)  = K e^{-au - ae^{-u}}
\end{equation}
with $u=b(y-s) $
 where $K,b,s$ are constants to be derived as functions of $a$. We write
 
 \begin{equation}
\eta = \ln a - b(y-s) = \ln a - u
\end{equation}

then $a e^{-u} = e^{\eta} $ and  $d \eta = -b dy$, and the $n^{th}$ moment
is given by

\begin{equation}
M_n=\int^\infty_{-\infty} y^n P(y)  dy = \frac{1}{b}\int^\infty_{-\infty} 
P(y) d\eta \frac{[\ln(a)+bs-\eta]^n}{b^n}
\end{equation}

Then, using A2, we write $P(y)$ (A1) as  
\begin{equation}
P(y) =  K \ e^{-a (\ln(a) - \eta) - e^{\eta}} 
           =  \bar{K} e^{a \eta - e^{\eta}}
\end{equation}
where $\bar{K}=Ke^{-a\ln(a)}$.

Now to within a constant we can write $M_n$ as:
\begin{equation}
\tilde{M}_n = \int_{-\infty}^{\infty} \eta^n P(y) d \eta
= \bar{K} \int_{-\infty}^{\infty} \eta^n e^{a \eta -e^{\eta}} d \eta
\end{equation}

so that $M_0=\tilde{M}_0/b$. Using the substitution
$\tau = e^{\eta}$ A5 becomes 
\begin{equation} 
\tilde{M}_n = \bar{K} \int_{0}^{\infty} (\ln \tau)^{n} \tau^{a-1} e^{-\tau} d \tau
= \bar{K} \frac{d^n}{da^n} \Gamma (a)
\end{equation}
where $\Gamma (a)$ is the Gamma function.
Thus 
\begin{eqnarray}
\tilde{M}_0 & = & \bar{K}\Gamma (a) \nonumber \\
\tilde{M}_1 & = & \bar{K}\Gamma (a) \Psi(a) = \tilde{M}_0 \Psi(a) \nonumber \\
\tilde{M}_2 & = & \bar{K}\Gamma (a) [ \Psi^2(a) + \Psi^{'}(a) ] \nonumber \\
              & = & \tilde{M}_0 ( \Psi^2(a) + \Psi^{'}(a) )
\end{eqnarray}
where
\begin{eqnarray}\nonumber
\Psi(a) = \frac{d\Gamma(a)}{da}\frac{1}{\Gamma(a)}. \\\nonumber
 \end{eqnarray}
 We now insist that $M_0=1$, $M_1=0$ and $M_2=1$.
 
 Thus
 \begin{equation}
 M_0 = \frac{\tilde{M_0}}{b} =\frac{\bar{K}\Gamma (a)}{b} =1
 \end{equation}
 and
 \begin{eqnarray}
 M_1 = & 0 & = \frac{1}{b^2}\int_{-\infty}^{\infty} P(y) d\eta [\ln(a)+bs-\eta] \nonumber \\
      &  & = \frac{1}{b^2}\left[ ( \ln (a) + bs) \tilde{M_0} -\tilde{M_1} \right]
 \end{eqnarray}
so 
\begin{equation}
\frac{\tilde{M_1}}{\tilde{M_0}}=\ln (a) + bs = \Psi (a)
\end{equation} 
from A7. Thus 
\begin{equation}
bs = \Psi (a) - \ln (a)
\end{equation}
Also 
\begin{eqnarray}
 M_2 = & 1 & = \frac{1}{b^3}\int_{-\infty}^{\infty} P(y) d\eta [\ln(a)+bs-\eta]^2 \nonumber \\
      &  & = \frac{1}{b^3}\left[ ( \ln (a) + bs)^2 \tilde{M_0} 
      -2 (\ln a + bs)\tilde{M_1} +\tilde{M_2}\right]
 \end{eqnarray}
 which, using A7 and A10 rearranges to give
\begin{equation}
M_2 = 1 = \frac{\tilde{M}_0}{b^3}\Psi^{'}(a).
\end{equation}
This finally gives the normalisation of the universal curve 
\begin{eqnarray}\nonumber
b^2= \Psi'(a)\nonumber\\
\bar{K}=\frac{b}{\Gamma(a)} \nonumber \\
{\rm that is  \ } K=\frac{b}{\Gamma(a)}e^{a \ln(a)}\label{gumnorm}\\
s=  \frac{(\Psi(a)-\ln(a))}{b}\nonumber
\end{eqnarray}
The above results will also yield an expression for the third moment in terms
of $a$. Following A3 and A5 we have 
 
\begin{eqnarray}
 M_3  & = & \frac{1}{b^4}\int_{-\infty}^{\infty} P(y) d\eta [\ln(a)+bs-\eta]^3 \nonumber \\
       & = & \frac{1}{b^4}\left[ ( \ln (a) + bs)^3 \tilde{M_0} 
      -3 (\ln a + bs)^2 \tilde{M_1} + 3 (\ln (a) + bs) \tilde{M}_2 - \tilde{M_3}\right]
 \end{eqnarray}
 Then A6 gives
 \begin{equation}
 \tilde{M_3} = \tilde{M_0} \left[\Psi (a) (\Psi^2 (a) + \Psi^{'}(a)) + 
 2 \Psi (a) \Psi^{'} (a) + \Psi^{''} (a)\right] 
 \end{equation}
which with A7 and A10 rearranges to give
\begin{equation}
M_3 = - \frac{\Psi^{''}(a)}{(\Psi^{'}(a))^{3/2}}
\end{equation}

\section{Moments of the Frechet distribution and normalization as a function of $a$.}

The moments of a Frechet distribution are obtained in \citet{bury}. Here we
wish to  consider PDF of the form (19) which has extremum statistics 
\begin{equation}
P_m(Q) = K (e^{u-e^u})^a
\end{equation}
where, following (25-32) we write:
\begin{equation}
u = \alpha + \beta \ln  (1 + \frac{Q}{\tilde{Q}})
\end{equation}
where here we use the notations 
$Q \equiv \Delta Q^{*}, \tilde{Q} \equiv \tilde{Q}^{*}$,
that is, $Q$ refers to extremal values. From (26), $\alpha$
and $\beta = (2k-1)$ are constants. We can then define
the moments of $P_m(Q)$:
\begin{equation}
M_n = \int_{-\tilde{Q}}^{\infty} \ Q^n \ dQ \ P_m (Q)
\end{equation}
since from B2 $u \rightarrow \infty$ as $ Q \rightarrow \infty$ and
$u \rightarrow -\infty$ as $ Q \rightarrow -\tilde{Q}$. Using the 
substitution $a e^{u} = \zeta$ we obtain after some algebra

\begin{equation}
M_n= \bar{K} \tilde{Q}^n \int_{0}^{\infty} ((\frac{\zeta}{g})^{1/\beta} - 1)^n
\ \zeta^{a-1+1/\beta} \ e^{-\zeta} \ d \zeta
\end{equation}
where the constants 
\begin{equation}
g=a e^{\alpha}\;\;\; {\rm  and}\;\;\; \bar{K} = \frac{K \tilde{Q}}
{\beta g^{\frac{1}{\beta}} a^a}.
\end{equation}
By taking the expansion $u=\alpha + \beta Q/\tilde{Q}$ it is
straightforward to verify that B4 yields the results from Appendix A.
We now insist that $M_0=1$,  $M_1=0$ and $M_2=1$.

B4 then gives 
\begin{equation}
M_0 = 1 = \bar{K} \Gamma (\bar{a}) \;\;\;{\rm \ where \ }\;\;\; \bar{a} = a + 1/\beta
\end{equation}
and
\begin{eqnarray*}
M_1 = 0 = \bar{K} \tilde{Q} [\frac{\Gamma (\bar{a} + 1/\beta)}{g^{1/\beta}}
- \Gamma (\bar{a})] \nonumber
\end{eqnarray*}
that is
\begin{equation}
\Gamma (\bar{a} + \frac{1}{\beta}) = g^{1/\beta} \Gamma (\bar{a})
\end{equation}
and using B7 we have from B4:
\begin{eqnarray*}
M_2 = 1 = \bar{K} \tilde{Q}^2 [ \frac{\Gamma^2 (\bar{a}) \Gamma (\bar{a} 
+ 2/\beta)}{\Gamma^2 (\bar{a} + 1/\beta)} - \Gamma (\bar{a})] \nonumber
\end{eqnarray*}
that is
\begin{equation}
 1 =  \tilde{Q}^2 [ \frac{\Gamma (\bar{a}) \Gamma (\bar{a} 
+ 2/\beta)}{\Gamma^2 (\bar{a} + 1/\beta)} -  1]
\end{equation}
using B6.

Now from the main text (27) $a = \frac{2k}{2k-1}$ and since
\begin{eqnarray}
\beta & = & - (2k-1) \nonumber \\
\bar{a} & = & a + 1/\beta  = 1 
\end{eqnarray}
and $\Gamma (\bar{a}) = \Gamma (1) = 1$.

B7 then gives $g^{1/\beta} = \Gamma (1+1/\beta)$. B8 then gives
$\tilde{Q}$:
\begin{equation}
\tilde{Q}=\pm\frac{\Gamma(1+\frac{1}{\beta})}{\left[\Gamma(1+\frac{2}{\beta})-\Gamma^2(1+\frac{1}{\beta})
\right]^\frac{1}{2}}
\end{equation}
then B7 gives K as
\begin{equation}
K=\pm \frac {\beta a^a \Gamma (1+1/\beta)}{\tilde{Q}}
\end{equation}
and since $g=a e^{\alpha}$, B6 gives an expression for $\alpha$:
\begin{equation}
(ae^\alpha)^{\frac{1}{\beta}}=\frac{K \tilde{Q}}{\beta a^a}
\end{equation}
that is:
\begin{equation}
\alpha=-\beta \ln\left(\frac{a^{\frac{1}{\beta}}}{\Gamma(1+1/\beta)}\right)
\end{equation}
which completes the normalization of
B1,B2 as  functions of $k$ or $a$.

Using B7 we have from B4 an expression for the third moment:
\begin{equation}
M_3 =\bar{K}\tilde{Q}^3 \left[ \frac{\Gamma(\bar{a}+\frac{3}{\beta})\Gamma^3 (\bar{a})}{\Gamma^3 
(\bar{a} + \frac{1}{\beta})}  - \frac{3\Gamma(\bar{a}+\frac{2}{\beta})\Gamma^2 (\bar{a})}{\Gamma^2 
(\bar{a} + \frac{1}{\beta})} + \frac{3 \Gamma(\bar{a}+\frac{1}{\beta})\Gamma (\bar{a})}{\Gamma
(\bar{a} + \frac{1}{\beta})} - \Gamma (\bar{a}) \right]
\end{equation}
Expansion in $1/\beta$ readily shows that to lowest order result A17 is recovered.

Then using B9, B10 and B11, B13 can be rearranged to give $M_3 (\beta)$, and hence $M_3$ as a 
function of $k$ or $a$:
\begin{equation}
M_3=\frac{\left[\Gamma(1+\frac{3}{\beta})-3\Gamma(1+\frac{2}{\beta})
\Gamma(1+\frac{1}{\beta})+2\Gamma^3(1+\frac{1}{\beta})\right]}
{\left[\Gamma(1+\frac{2}{\beta})-\Gamma^2(1+\frac{1}{\beta})\right]^\frac{3}{2}}
\nonumber
\end{equation}
 
\bibliographystyle{egs}

\end{document}